\documentstyle[11pt]{article}
\begin{document}
Kolmogorov Complexity, Cosmic Microwave Background Maps and the
Curvature of the Universe
\vspace{0.2in}

V.G.Gurzadyan
\vspace{0.1in}

ICRA, Dipartimento di Fisica, Universita di Roma La Sapienza,
Roma, Italy; 
Yerevan Physics Institute and Garni Space Astronomy Institute, 
Armenia\footnote{E-mail: gurzadya@icra.it}

\vspace{0.2in}

{\bf Abstract.}
The information theory approach is suggested to the
Cosmic Microwave Background (CMB) problem for negatively curved
homogeneous and isotropic Universe.
Namely, the Kolmogorov complexity of anisotropy of spots in CMB
sky maps is proposed as
a new descriptor for revealing crucial cosmological information,
particularly on the curvature of the Universe.
Such profound descriptor can be especially valuable while analyzing the
data of forthcoming space and ground based experiments
MAP, Planck surveyor, CAT, etc.
\vspace{0.1in}


\vspace{0.2in}

The Cosmic Microwave Background
radiation data and particularly the sky maps are an
essential source of cosmological
information. Various descriptors have been
proposed to extract that information, including the
hot spot number density, genus, correlation function of local
maxima, Euler-Poincare characteristic, etc. (for review see \cite{Silk}).
Most of these descriptors have been already applied for the analysis of
COBE sky maps \cite{smoot95}. However, in view of forthcoming new
generation high precision
experiments, the importance of involving of
more refined descriptors, is evident.

In the present paper we turn attention to a new descriptor,
namely to {\it the Kolmogorov complexity}
of the CMB anisotropy spots, which can carry important
information on the geometry of the Universe. The idea is based on
the effect of geodesic mixing \cite{GK} occurring in
Friedmann-Robertson-Walker (FRW) Universe with
negative curvature. Several observable consequences of this effect
for CMB properties have been predicted
\cite{mixing},\cite{book}, including:
(a) decrease of the amplitude of the anisotropy after
the last scattering epoch, (b) flattening of the angular autocorrelation
function, (c) distortion of the sky maps.

Particularly, the distortion of anisotropy spots is a result of
strong statistical properties - exponential instability of geodesic flows in
space (locally if the space is non-compact) with constant negative
curvature  - Anosov systems \cite{An}. Namely, an exponential elongation
(stretching)
of phase space density occurs uniformly for the dimensions of the phase
space, including the spatial coordinates and, in accord to Liouville theorem
it contracts in equal number of dimensions.

The signature of this predicted effect - a threshold independent
elongation of spots has been discovered \cite{GT}
at the refined analysis of COBE 4 year (complete) dataset.
Given the absence of other reasonable physical
mechanisms, this can be a {\it direct indication of the negative curvature of
the Universe}.

The essence of the effect of geodesic mixing includes the projection
of geodesics from d=(3+1) pseudo-Riemannian space to d=3
Riemannian space and the study of the behavior of time correlation
functions of geodesic flows in 3-spaces.

The problem of the geodesics' projection for homogeneous metrics was
studied by Lockhart, Misra and Prigogine \cite{LMP}; in that case the
reparametrization of the time of the geodesic is as follows
$$
\lambda(t)= \int_{t_0}^{t} a^{-1}(s)ds = \eta(t) - \eta(t_0).
$$
At d=3 the time correlation function of geodesic flow
$f^t$ is decreasing by exponential law,
i.e. $ \exists c>0$ such that for all
$A_1, A_2 \in L^2(SM)$ \cite{Pol}
$$
\bar b(t) = \mid \int A_1 (f^tu) A_2(u)d\mu - \int A_1(u)d\mu(u)
\int A_2(u)d\mu(u)\mid=
$$
\begin{equation}
c \parallel A_1\parallel \parallel A_2 \parallel
\dot (1+t) e^{-h(f)t} + O(e^{-ht}).
\end{equation}
where $\mu(SM)=1$ is the Liouville measure and
$$
\parallel A \parallel = [\int A(u)^2 d\mu(u)]^{1/2},
$$
and $h$ is the Kolmogorov-Sinai (KS) entropy.

This exponential law is determining the quantitative efficiency of geodesic
mixing, i.e. of the 3 observable effects mentioned above, including the
specific elongated distortion of CMB maps\footnote{Note, that the 
map distortion due to geodesic mixing has no direct relation with the
effects discussed in \cite{Bar}, and exists whatever are the initial 
shapes of spots at the last scattering surface.}.
However, such elongation is a simplification of
the complexity of the anisotropy spots, as mentioned already in \cite{book}.
To describe quantitatively the latter here we suggest to use the
invariant definition of complexity -
Kolmogorov complexity - introduced by Kolmogorov in 1965
\cite{Kolm}. The fundamental consequences of the concepts of complexity
and random sequences were revealed in studies by Solomonoff, Martin-Lof and
especially, by Chaitin (see \cite{Ch}) and concerning the basics of physics,
e.g. second law of thermodynamics, by Zurek \cite{Zu}.

Kolmogorov complexity $K_u$ is defined as the minimal length of the
binary coded (in bits) program which is required to describe the system
completely \cite{ZL}, i.e. it will enable to recover the initial system via
a given computer. The Kolmogorov complexity of object $y$ at
given object $x$ is defined as
\begin{equation}
K_{\phi(p)}(x) =min_{p:\phi=x} l(p),
\end{equation}
where $l(p)$ is the length of the program $p$ with respect to
computer $\phi(p)$
describing the object completely, i.e. at $0-1$ representation:
$l(\O)=0$. The fundamental point of the Kolmogorov's formulation is the
independence of complexity on the computer. Therefore the Turing machine
can be considered as a universal computer while computing the complexity.

Thus, the Kolmogorov complexity is the amount of information which is
required to determine uniquely the object $x$. Kolmogorov had proved that
the amount of information of $x$ with respect to $y$ is given by the
formula \cite{Kolm}
$$
I(y:x)=K(x)- K(x\mid y),
$$
where the conditional complexity
$$
K(x\mid y)=min\, l(p),
$$
is the minimal length of the program required to describe the object $x$ when
the shortest program for $y$ is known.
Note, that
$K(x\mid x)=0$ and $I(x:x)=K(x)$ and $I(x:x)\succeq 0$;
where $A \succeq B$ denotes $A\leq B + const$.

Kolmogorov complexity measured in bits is related to KS-entropy via
the relation \cite{Cav}
\begin{equation}
\Delta I =log_2 (2^{h(f^t)(t-t_0})=h(f^t)(t-t_0),
\end{equation}
where the loss of information $\Delta I$ during the time interval $t-t_0$ is
\begin{equation}
\Delta I = K_u(t) - K_u(t_0),
\end{equation}
i.e. the information corresponding to the distortion of the pattern from
the state initial state $t_0$ (the last scattering epoch) up to the observer
at $t$ i.e. at $z=0$.
This is guaranteed by Shennon-McMillan-Breiman theorem \cite{Shen} stating the
uniform exponential rate of loss of information.

Return again to the cosmological problem. In $k=-1$
FRW Universe any initial CMB structure observed
at redshift $z_{obs}$ should have more complex (amoebae-like) shape
than at $z < z_{obs}$
$$
K(CMB spot\mid z=z_{obs}) > K(circle\mid z< z_{obs});
$$
here 'circle' denotes a circle with Gaussian or other fluctuations.
The complexity estimated for the anisotropy spot observed now
should exceed the complexity of the primordial spot (besides
the scale expansion $1+z$ times)
by an amount depending on the KS-entropy and, hence on the curvature
and the distance of the last scattering surface. Indeed,
for the last scattering epoch at redshift $z$ for the Universe with the
present density parameter $\Omega_0$ the exponential factor in (1) yields
\cite{mixing},\cite{book}
\begin{equation}
e^{ht}= (1+z)^2[1+\sqrt{(1-\Omega_0})/(\sqrt{1+z\Omega_0} +
\sqrt{1-\Omega_0})]^4
\end{equation}
For FRW Universe
KS-entropy is determined only by the scale factor of the Universe
(this is natural, since no other scale factor exists in FRW spaces) and
is equal \cite{GK}
\begin{equation}
h=2/a,
\end{equation}
Then, from Eqs.(3) and (4) we come to a simple equation
linking the geometry with the complexity
\begin{equation}
\Delta K = (2/a) \Delta T.
\end{equation}
In other words, the relative complexity of the observed spot with respect
to a circular spot (which one can expect for flat or positively curved
spaces) will
determine the curvature of the hyperbolic Universe, where $\Delta T$ is the
time elapsed since photons started to move freely and thus
tracing how curved the 3D space is.

Kolmogorov complexity $K$, therefore, enables one to reduce the properties of
CMB and the geometry of the Universe to a simple pattern recognition
analysis of the shapes of anisotropy spots. Namely, the numerical value of
$K$ estimated by means of the analysis of CMB sky maps will carry information
on the curvature, density parameter $\Omega$, the redshift of the last
scattering epoch and the law of expansion of the Universe in post-scattering
epoch.

The cosmological information should be extracted from the CMB maps in the
following way. First, certain criterion should define the spots represented via
given configurations of pixels (e.g.\cite{GT}).
Then, the computation of the Kolmogorov
complexity should be performed by means of the length of a special compressed
code (string) completely defining the spots. Since the basic program will
be the same,
the only changes will be due to different data files, i.e. the coordinates of
the pixels of various spots. This problem technically has been solved in
\cite{AGS} and the Kolmogorov complexity of given test configurations has been
computed along with their Hausdorff dimension. Note, that COBE-DMR data are
still not efficient for calculation of such a refined descriptor as the
complexity, since have the ratio
signal/noise=2/1; the latter enables to determine certain genuine spots from
the noise
\cite{Tor}\cite{Cay} but not the reliable shape of the spots. However, the
forthcoming experiments, such as MAP and Planck Surveyor,
can well enable to define the anisotropies
with sufficient accuracy, and hence, to elaborate the technique
developed here.

Thus, the heuristic content of this approach is clear:

(a) CMB properties  are reduced to a problem of random
sequences and information theory known by its fundamental 
achievements \cite{Ch};

(b) Kolmogorov complexity of CMB anisotropies is a
computable descriptor containing direct information on the geometry of
Universe.

Moreover, the CMB properties discussed
above, can be only one of manifestations of a much deeper link -
{\it negatively curved space - mixing - complexity} with the fundamental
physical laws, namely, implying {\it mixing - second law of thermodynamics 
- arrow of time}. Thus, {\it we observe the second law and the time asymmetry 
because 
we live in a Universe with negative curvature, and those laws may not
be the same in a flat
or positively curved spaces}. We plan to discuss curvature/second-law 
conjecture, including the
relations between thermodynamic, cosmological and other arrows of time, 
in a separate paper.  

The complexity and information way of thinking can be valuable also
in other key cosmological problems. 

I am thankful to W.Zurek and my colaborator  A.Allahverdyan for 
valuable discussions and the referee for helpful comments.

\end{document}